\documentclass[useAMS,usenatbib]{mn2e}
\usepackage{graphicx,graphics}

\title[Radio polarization measurements from RRAT J1819$-$1458]{Radio polarization measurements from RRAT J1819$-$1458}
\author[Karastergiou et al.]
{A.~Karastergiou$^1$, A.~W~Hotan$^2$, W. van Straten$^3$, M.~A~McLaughlin$^{4,5,6}$ and S.~M~Ord$^7$\\
$^1$Astrophysics, University of Oxford, Denys Wilkinson Building, Keble
  Road,
  Oxford OX1 3RH, UK\\
$^2$Curtin Institute of Radio Astronomy, Curtin University of
Technology, GPO Box U1987 Perth, Western Australia, 6845\\
$^3$Swinburne University of Technology, Centre for Astrophysics and
Supercomputing, Mail 31, PO Box 218, VIC 3122, Australia\\
$^4$Department of Physics, West Virginia University, Morgantown, WV
26506, USA\\
$^5$National Radio Astronomy Observatory, Green Bank, WV 24944\\
$^6$Alfred P. Sloan Foundation Fellow\\
$^7$Harvard-Smithsonian Center for Astrophysics, 60 Garden St, Cambridge, MA, USA
}
\date{\today}
\begin{document}

\date{started 18 November}

\pagerange{\pageref{firstpage}--\pageref{lastpage}} \pubyear{2008}

\maketitle

\label{firstpage}

\begin{abstract}
  We present the first polarization measurements of the radio emission
  from RRAT J1819$-$1458. Our observations, conducted in parallel to
  regular timing sessions, have yielded a small number of bright and
  polarized pulses. The polarization characteristics and integrated
  profile resemble those of normal pulsars with average spin-down
  energy $\dot E$: moderate to low linear polarization in the
  integrated profile despite relatively high polarization in the
  individual pulses.  On average, a small degree of circular
  polarization is also observed. The polarization position angle
  executes a remarkably smooth, steep S-shaped curve, interrupted by
  two orthogonal jumps. Based on the shape of the PA swing, we place
  some constraints on the emission geometry. We compare these
  polarization properties to those of other radio emitting neutron
  star populations, including young pulsars, pulsars with a high
  surface magnetic field and radio emitting magnetars.  From the
  polarization measurements, the Faraday rotation measure of this RRAT
  is derived.
\end{abstract}


\section{Introduction}

In the last few years, the family of radio emitting neutron stars has
been extended by the discovery of radio emission from magnetars
(Camilo et al. 2006)\nocite{crh+06}, intermittent pulsars (Kramer et
al. 2006)\nocite{klo+06}, and extremely sporadic but periodic pulses
from Rotating Radio Transients (RRATs, McLaughlin et
al. 2006)\nocite{mll+06}. The physical processes that lead to the
observed properties of these new types of astrophysical objects are
not well understood despite their classification as neutron
stars. This is not surprising given the open problem of radio emission
in normal pulsars.  Detailed studies of the radio emission properties,
however, not only reveal the links between these extreme objects and
normal pulsars, but also provide additional clues regarding the
mechanism of pulsar radio emission.

RRAT J1819$-$1458 has a rotational period of $P\approx$4.26~s, and a
relatively high frequency of events compared to the other
RRATs. McLaughlin et al. (2006) report a rate of roughly 1 pulse
detection every 3 minutes, while Esamdin et al. (2008)\nocite{ezy08}
report a detection of 162 bright ($>5\sigma$) pulses in a total of 94
hours of observation.  The period $P$ and period derivative $\dot P$
place this RRAT in a sparsely populated region of the $P$-$\dot P$
diagram, between the main pulsar population and magnetars. Given $P$
and $\dot P$, the estimated surface magnetic field is $B_s\approx 4.9
\times 10^{13}$ Gauss, the characteristic age is a relatively young
$\tau_c\approx 119.8$ kyr and the rate of rotational energy loss is a
relatively average $\dot{E}\approx 2.88 \times
10^{32}$erg~s$^{-1}$. Clearly RRAT J1819$-$1458 is not a typical
pulsar. In addition to the sporadic emission properties, the
literature reveals that sources with a relatively young characteristic
age usually feature high $\dot E$ values. It can therefore be used to
test whether $\dot E$ rather than age determines the degree of
polarization as shown by various studies (e.g. von Hoensbroech et
al. 1998; Johnston et al. 2006; Weltevrede \& Johnston
2008)\nocite{hkk98,jkw06,wj08b}.  Observational evidence and
theoretical considerations suggest that the degree of linear
polarization is regulated by the relative intensity of competing modes
of propagation in the pulsar magnetosphere, orthogonal in polarization
(e.g. Stinebring et al. 1984, Karastergiou et
al. 2002)\nocite{scr+84}. In many young and highly energetic pulsars,
only a single mode is observed. The age and $\dot E$ of pulsars have
also been associated with the characteristics of any observed high
energy emission. More specifically, young pulsars show a more
pronounced non-thermal component in their X-ray spectra than older
pulsars, whereas high $\dot E$ pulsars are very often $\gamma$-ray
emitters (Thompson 2004)\nocite{tho04}. RRAT J1819$-$1458 shows
thermal X-ray emission, entirely in-line with what is observed in
other pulsars of this characteristic age (McLaughlin et
al. 2007).\nocite{mrg+07}

On the other hand, it is interesting to seek comparisons in the
polarization characteristics between RRAT J1819$-$1458 and other
sources with a high inferred surface magnetic field, the most extreme
being the two known radio-emitting magnetars. Unfortunately, there are
to date no polarization measurements of the highest $B_s$ pulsars so
only a limited comparison is possible with slightly lower $B_s$
pulsars, which we carry out in Section 3 of this paper. Polarization
measurements of the two radio-emitting magnetars exist and show very
high linear polarization. Both XTE J1810$-$197 (Camilo et al. 2007;
Kramer et al. 2007)\nocite{crj+07,ksj+07} and 1E~1547.0$-$5408 (Camilo
et al. 2008)\nocite{crj+08} show almost 100\% linear polarization in
their profiles, the latter being the case at frequencies above
3~GHz. At 1.4~GHz, the polarization of 1E~1547.0$-$5408 drops to
around 25\%, although the reasons appear more likely related to
interstellar scattering than intrinsic, which would lead to
depolarization toward higher rather than lower frequencies
(e.g. Karastergiou et al. 2002)\nocite{kkj+02}. The polarization
position angle (PA) tentatively traces a dipolar magnetic field,
resulting in the well known S-shaped curve of the single rotating
vector model (hereafter RVM) of Radhakrishnan \& Cooke
(1969)\nocite{rc69b}, but there is also evidence that the PA swing is
variable with time. The radio emission mechanism of these sources is
not well understood and it exhibits an unconventional flat spectrum
across a very broad range of frequencies.
\begin{figure} 
\includegraphics[width=0.99\columnwidth]{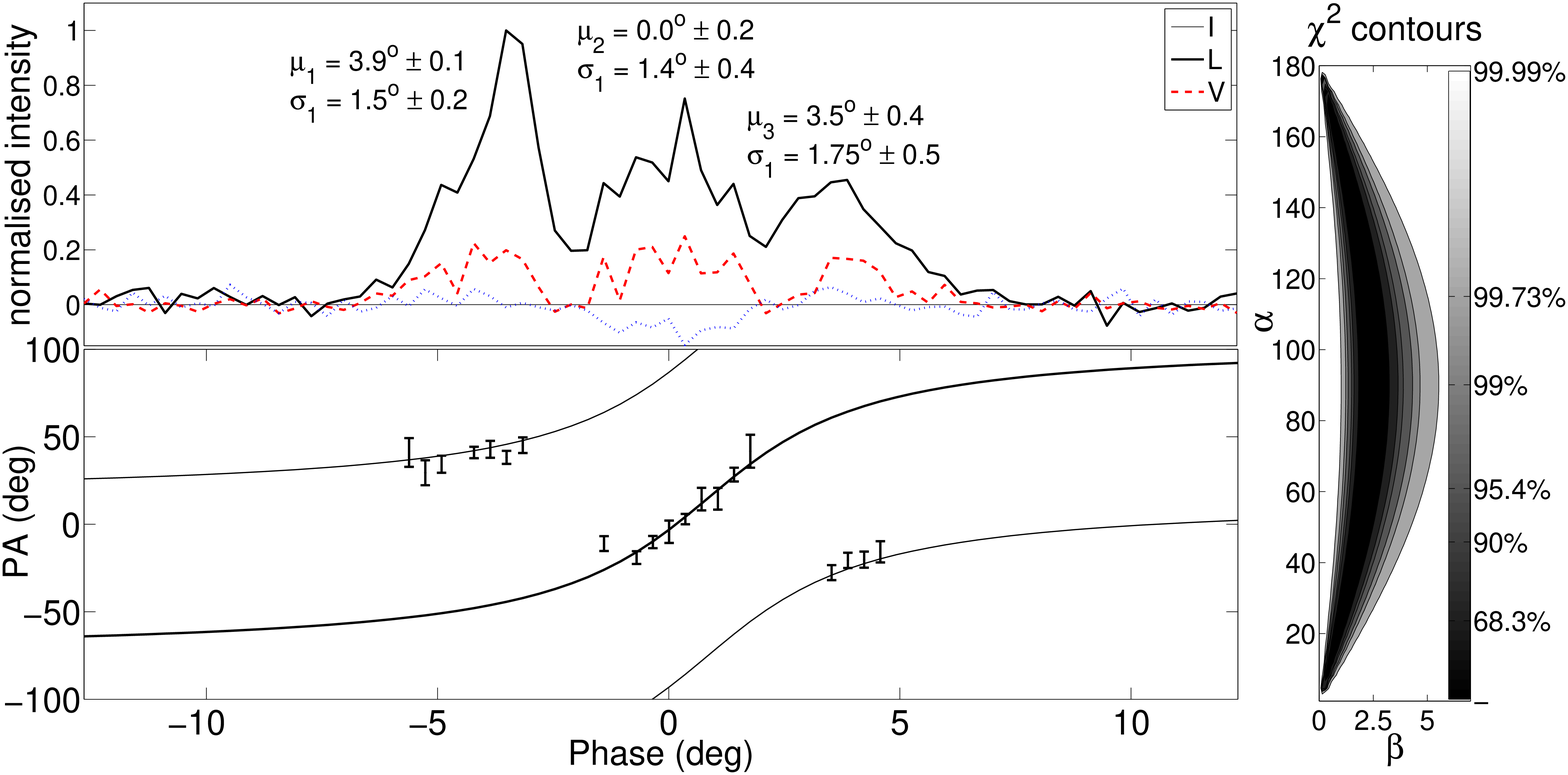}
\caption{The integrated polarization profile of RRAT J1819$-$1458 at
  1405 MHz. The top panel shows the 3-component profile made up of 72
  bright pulses, with the linear and circular polarization in dashed
  red and dotted blue line. The mean ($\mu$) and standard deviation
  ($\sigma$) of a 3-component Gaussian fit is shown for each
  component. The bottom panel shows the polarization PA (where linear
  polarization is greater than 2.5 $\sigma$), and a fit to the
  rotating vector model ($\alpha=96\degr$ and $\beta=2.57\degr$); the
  leading and trailing parts are offset by 90$\degr$ to the PA of the
  central pulse component. The $\chi^2$ probability distribution for
  $\alpha$ and $\beta$ on the right, shows the most likely parameter
  values lie within the dark, banana-shaped region. \label{avprof}}
\end{figure}

Weltevrede et al. (2006)\nocite{wsrw06} identified common elements of
RRAT J1819$-$1458 to the anomalous amplitude distribution of the
unusual pulsar B0656+14. This pulsar sits quite central in the bulk of
normal, middle aged pulsars on the $P$-$\dot P$ diagram, and is a
source of pulsed, high-energy emission. Similar to highly energetic
pulsars (for PSR B0656+14, $\dot E=3.8 \times 10^{34}$~erg s$^{-1}$)
the polarization of this pulsar is very high (Gould \& Lyne
1998)\nocite{gl98} at 1.4 GHz. The PA also shows smooth variation in
accordance with the RVM (Lyne \& Manchester
1988)\nocite{lm88}. Interestingly, an observation at 4.85~GHz (von
Hoensbroech 1999)\nocite{hoe99} shows reduced linear polarization and
orthogonal PA jumps in the profile, whereas Johnston et
al. (2006)\nocite{jkw06} show a total absence of polarization at
8.4~GHz, and discuss the peculiar and abrupt depolarization of this
young pulsar towards high frequencies.

In the following we show the first results from polarization
measurements of RRAT J1819$-$1458, in an attempt to identify
similarities with the objects mentioned above and to understand the
emission geometry of this RRAT. The observations are described and the
results of our analysis are put forward. A brief discussion based on
the results and possible interpretations ensues, followed by
concluding remarks.  \vspace*{-0.5cm}
\section{Observations and data analysis}
Data were obtained at the Parkes radio telescope on 13 epochs from May
2007 to September 2008, using the central beam of the Parkes 21cm
multi-beam receiver and the Caltech Parkes Swinburne Recorder Mk II
(CPSR2); (Bailes 2003)\nocite{bai03}. Two 64 MHz-wide
dual-polarisation bands centred on 1341 and 1405 MHz were sampled at
the Nyquist rate with 2-bit precision and the raw data recorded to
disk for off-line processing. Techniques similar to those described by
Knight et al. (2005)\nocite{kbmo05} were used to search roughly 0.5~h
of data per epoch for individual, high signal-to-noise
pulses. Dedispersion, calibration and further processing were
performed using standard pulsar software packages, DSPSR and PSRCHIVE
(Hotan et al. 2004)\nocite{hvm04}. The data were smoothed to 8192 bins
across the profile, achieving a temporal resolution of roughly
0.52~ms. Each of the 64-MHz sidebands was split into 32 frequency
channels.

The two CPSR2 sidebands were calibrated independently, and inspected
closely for individual pulses. The automatic flagging yielded roughly
20 pulse candidates per epoch, at the $\approx$4.26~s period. These
were then carefully examined by eye, to reject spurious signals of
interference and only pulses of significant S/N after dedispersion
that appeared in both CPSR2 sidebands at the highest temporal
resolution were considered. As a consequence, an average of $\approx$6
pulses were kept per epoch. This extremely cautious procedure resulted
in a rate of approximately 10 pulses per hour, which lies between the
reported McLaughlin et al. (2006) and Esamdin et al.(2008) rates. It
also ensured the total 72 pulses used in this analysis to be bright
and contain reliable polarization information. A coherent timing
solution was built using these very narrow pulses, in order to obtain
the phase connection between the epochs.  The small total number of
pulses does not permit a statistical analysis on the variability of
the polarization, but forms an excellent sample to address the first
order polarization issues.
\section{Results}
\begin{figure} 
\includegraphics[width=0.49\textwidth]{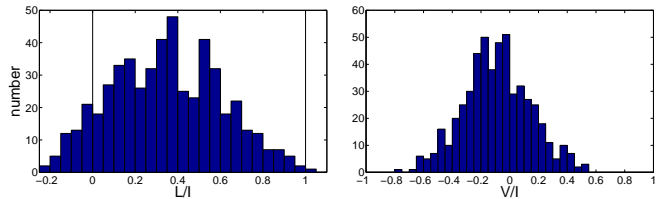}
\caption{Histograms of the fractional linear polarization L/I (left)
  and fractional circular polarization V/I (right). Only bins where
  the total power S/N is greater than 5 are considered. On the left,
  values above 1 and below zero are due to instrumental noise. The
  histograms have means of 37\% for L/I and $-$6\% for
  V/I. \label{overi}}
\end{figure}
\subsection{Faraday Rotation Measure}
With a dispersion measure (DM) of 196 cm$^{-3}$pc, it is expected that
Faraday rotation will effectively reduce the degree of
polarization. The Rotation Measure (RM) was estimated in two ways. The
first was to trace the PA of highly polarized pulses across each of
the two CPSR2 sidebands, and fit the PA for the well-known dependence
on the wavelength squared ($\lambda^2$). The second was to apply a
range of RM values to each pulse, then collapse the data in frequency
and trace how the degree of linear polarization changes with RM. The
correct RM recovers the maximum linear polarization. With the small
number of pulses, the error in the RM measurement is non-negligible,
however both methods result in similar values of $\approx$330$\pm$30
rad/m$^2$. This value corresponds to an average parallel magnetic
field component of $<B||>\approx$2$\mu$Gauss, which is somewhat larger
than expected at the nominal distance of roughly 3.5 kpc (according to
the NE2001 model of Galactic free electron density of Cordes \& Lazio,
2002\nocite{cl02}), although of the expected sign for this rather
poorly sampled line of sight. This RM was used to correct the data and
regain the maximum polarization.
\subsection{Integrated pulse profile}
Polarization is an invaluable diagnostic of the radio emission, and
even without complete knowledge of the physical processes involved,
the polarization can help answer a number of questions. The first
question relates to the emission geometry. According to the rotating
vector model for pulsar polarization, the observed PA is tied to the
magnetic field lines, changing smoothly as the line-of-sight
intersects different field lines at different angles, as:
\begin{equation}
\tan{(PA-PA_0)}=\frac{\sin({\phi-\phi_0})\sin{\alpha}}{\sin{\zeta}\cos{\alpha}-\cos{\zeta}\sin{\alpha}\cos({\phi-\phi_0})}.\label{rvm}
\end{equation}
PA$_0$ and $\phi_0$ are constant offsets in PA and phase, $\phi$ the
pulse phase, $\alpha$ the inclination angle between magnetic and
rotation axis, and $\zeta$ the sum of $\alpha$ and the impact
parameter $\beta$, which is the nearest angle of approach of the line
of sight to the magnetic axis. This equation is derived with the
convention that the position angle increases clockwise on the sky.
Figure \ref{avprof} shows the integrated polarization profile from 72
pulses of RRAT J1819$-$1458, binned to a temporal resolution of
4.16~ms (1024 bins across the profile), with the total power (solid),
linear (dashed) and circular polarization (dotted) in the top
panel. The bottom panel shows the measured PAs where the linear
polarization is greater than 2.5~$\sigma$.  The line represents a fit
to Eq. \ref{rvm}, taking into consideration possible orthogonal
jumps. The parameters of the fit shown are ${\rm PA}_0=15\degr$ (at
the frequency of observation), $\phi_0=0.82\degr$, $\alpha=96\degr$
and $\beta=2.57\degr$.  Fitting of $\alpha$ and $\beta$ was performed
via a ``brute-force'' method of scanning across the 2-D parameter
space. The reduced $\chi^2$ probability contour map of $\alpha$,
$\beta$ pairs, shown on the right, indicates a very high probability
that the solution lies within the shaded, banana-shaped region; the
solution for $\beta$ is well constrained, while quite the opposite is
true for $\alpha$.

Fig. \ref{avprof} paints a familiar picture reminiscent of many normal
pulsars, with a 3-component structure of the integrated pulse. As an
integration of only 72 pulses, it is merely indicative of the average
profile. This imposes limitations on the interpretation of the profile
components and widths, which we obtained by fitting (3) Gaussian
components, the mean $\mu$ and standard deviation $\sigma$ of which
are indicated on Fig. \ref{avprof}. These measurements provide two
additional, model-dependent, methods to cautiously constrain the
geometry, which are outlined in Section 1.3 of Everett \& Weisberg
(2001)\nocite{ew01} and both make use of the value of the steepest PA
gradient, here $\approx$24$\degr$/degree.  The first originates from
the patchy beam model of Lyne \& Manchester (1988) and also relies on
a measurement of the profile width at 10\% maximum, assuming emission
from both sides of a circular beam. Solving Eqs. 5, 6 and 7 of Everett
\& Weisberg for a 10\% width of $11.2\degr$ results in
$\alpha=40\degr$, $\beta=1.6\degr$. The second originates from Rankin
(1990)\nocite{ran90}, which assumes the central component represents
``core'' emission and depends on the period and $\alpha$ as Eq. 9 of
Everett and Weisberg. This yields a solution of $\alpha=21\degr$,
$\beta=0.9\degr$, assuming a half maximum width of the central
component of 3.3$\degr$.  Both calculations are in agreement with the
contour map in Fig.~1.

The 72 individual pulses, mostly consist of a single, very narrow
pulse that falls under one or the other component each carrying its
own polarization signature. The three component profile is then
consistent with the S-shaped PA swing across the entire profile. The
timing solution used here is independent of the polarization. The fact
that it yields an integrated PA swing which very closely follows the
RVM curve lends significant weight to this solution and hence to the
integrated profile presented here. Furthermore, this clearly answers
the question regarding the reported 2 bands of timing residuals in
Esamdin et al. The bands are in fact 3, owing to the 3 components of
the profile and the single Gaussian pulse used for the timing template
matching.

Focusing on the polarization, at least 2 clear orthogonal jumps are
present in the integrated PA swing, around pulse phase -3.5$\degr$ and
+3$\degr$ from the designated origin. It is no surprise then that the
total degree of linear polarization is relatively low (20-30\% under
each component), as a result of competing orthogonal modes of
polarization. The circular polarization (V) is very low, there is
however a tentative sense reversal at a pulse phase of
$\approx$1$\degr$. The pulse phase of the steepest PA gradient is
poorly defined, as the gradient is approximately constant in the
middle component, but it appears to slightly lag the component mean,
as would be the case due to kinematic effects of the pulsar rotation
(e.g. Blaskiewicz et al. 1991)\nocite{bcw91}. More exact pulse phase
locations for the mean of the central component, the steepest PA
gradient and the sense reversal in $V$ will provide an estimate of the
emission height when a true average profile becomes available.
\subsection{Individual pulses}
\begin{figure*} 
\includegraphics[width=1.\columnwidth]{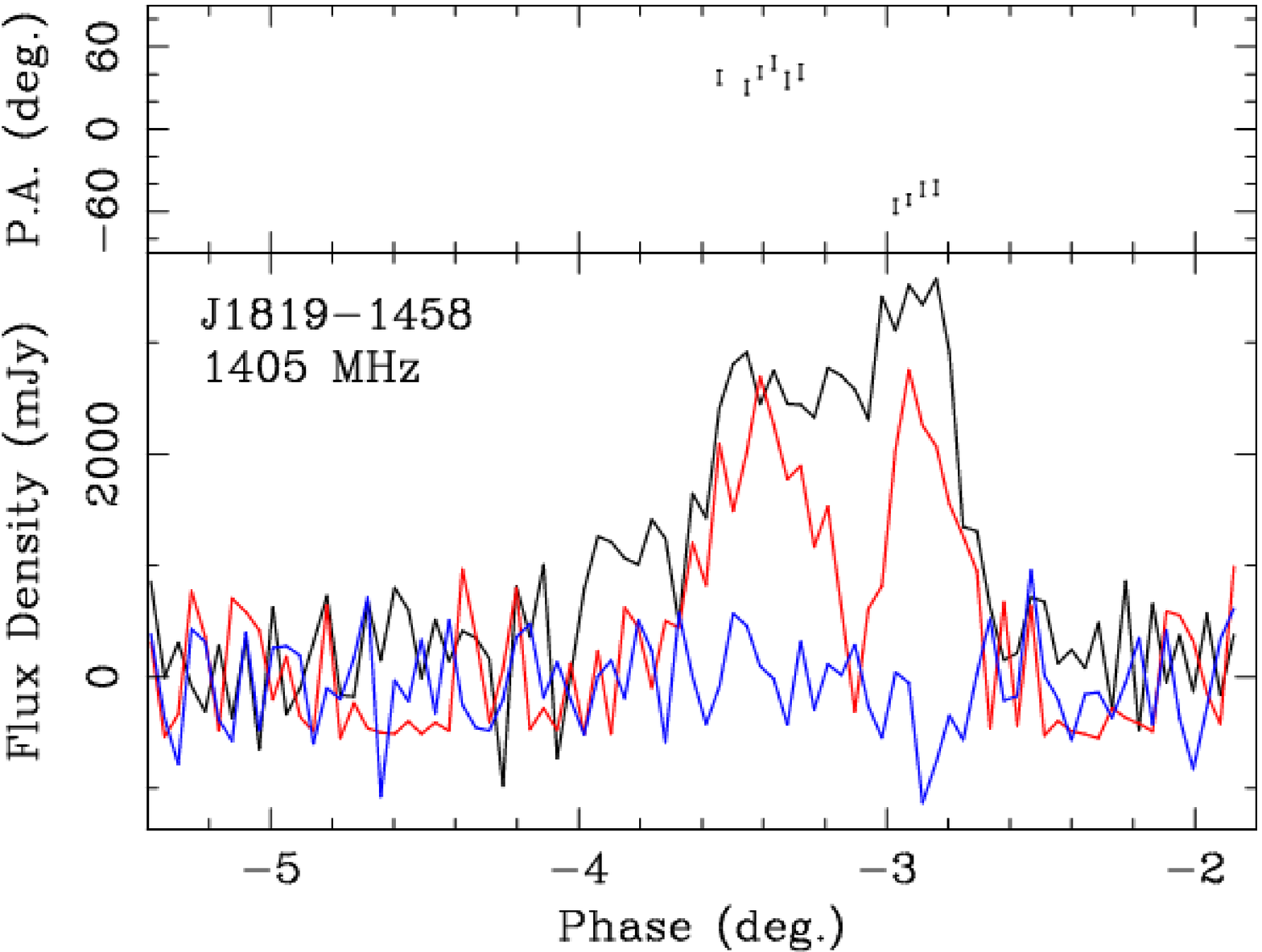}
\includegraphics[width=1.\columnwidth]{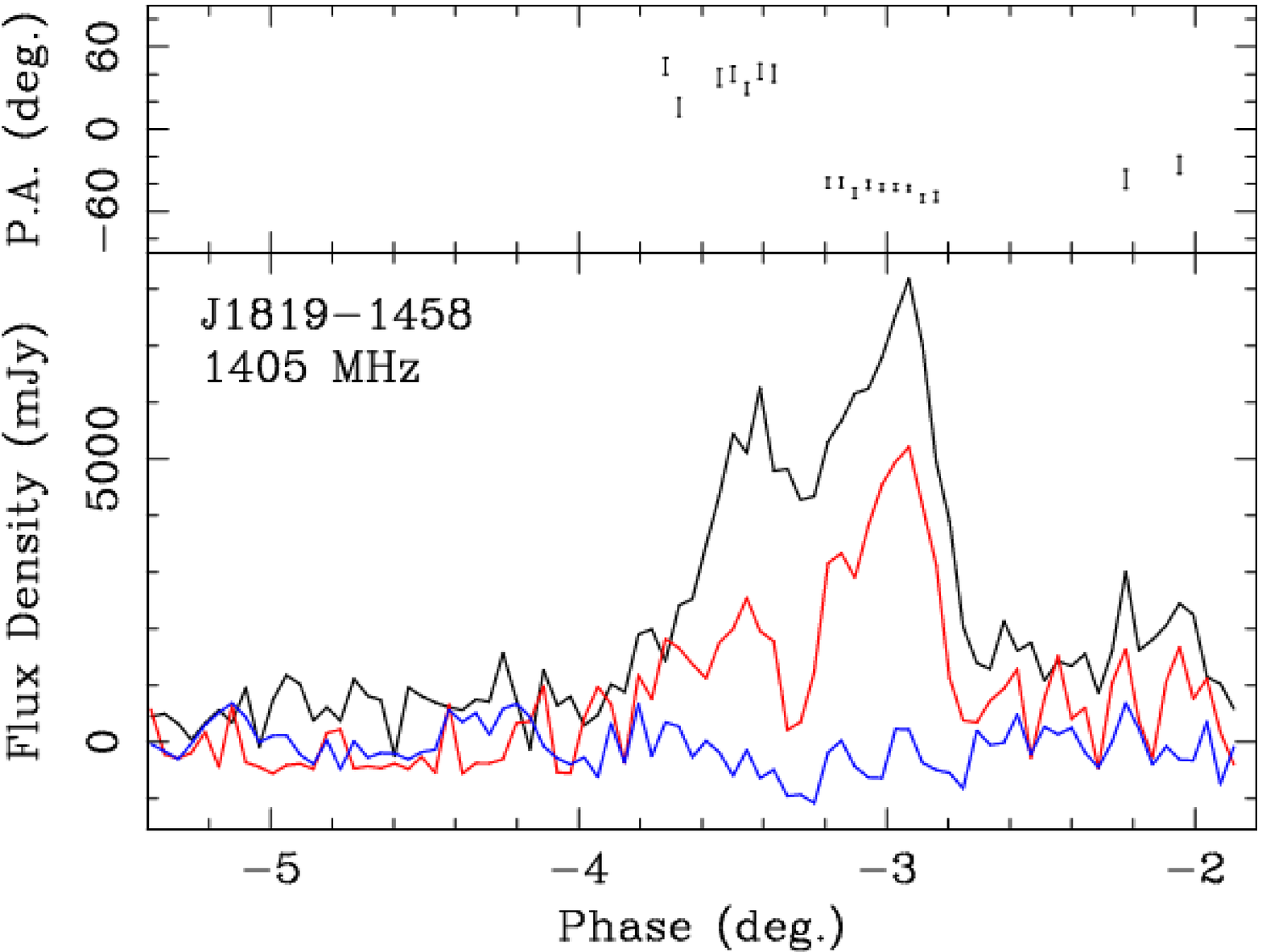}
\caption{Two pulses that show an orthogonal jump in the PA (top panel)
  coinciding with a sudden drop in the linear polarization (dashed red
  line, bottom panel). The total power is in solid black line and the
  circular polarization in dotted blue line. Both pulses are located
  towards the leading edge of the integrated profile. \label{opm}}
\end{figure*}
A closer look at the individual pulses that make up the profile of
RRAT J1819$-$1458 reveals narrow pulses often comprising of a small number
of sub-pulses, a few ms in width, with variable degrees of linear and
circular polarization, as seen in normal pulsars. Figure \ref{overi}
(left) shows a histogram of the degree of linear polarization in
individual bright data bins (in total power) with S/N$>$5. The degree
of linear polarization ranges from 0 to 100\% (the values on either
side of these limits are due to instrumental noise), with a mean of
37\%. This is higher than the linear polarization in the integrated
pulse profile, where the addition of pulses in both orthogonal modes
reduces the polarization.  Figure \ref{overi} (right) shows the
histogram of the degree of circular polarization, for the same high
S/N data bins. The histogram is slightly asymmetric and centred on
$-$6\%. In a small number of bins, the degree of circular polarization
exceeds 50\% in either direction.  In the pulses that show clear
evidence of 90$\degr$ PA jumps, the linear polarization drops to a
sudden minimum at the pulse phase of the jump, as expected by the
superposition of orthogonal polarization modes of equal intensity. Two
examples are shown in Figure \ref{opm}, where the pulse phase
reference is the same as Figure \ref{avprof}.

\section{Discussion}
The polarization observations presented here, albeit from a small
sample of pulses, are very revealing as to the nature of the radio
emission from RRAT J1819$-$1458. With low linear polarization and an
average value of $\dot E$, RRAT J1819$-$1458 clearly follows the
observational paradigm of normal pulsars, also providing further
evidence of the link between $\dot E$ and the degree of polarization.
Furthermore, as RRAT J1819$-$1458 is a relatively young pulsar, this
work again demonstrates that $\dot E$ rather than age is the critical
parameter as concerns the degree of linear polarization. The
orthogonal PA jumps seen at this frequency are also more typical of
pulsars with average $\dot E$ than of young pulsars. They also totally
distinguish this source from the two radio emitting magnetars, and PSR
B0656+14 at this frequency.

At present it is difficult to investigate a possible link between very
high surface magnetic field and degree of polarization, owing to a
lack of polarization measurements of the highest $B_s$ pulsars. There
are, however, some high quality polarization profiles from a number of
pulsars with $B_s>10^{13}$. The parameters and polarization
characteristics of these profiles, found in Gould \& Lyne (1998),
D'Allessandro \& McCulloch (1997)\nocite{dm97}, Qiao et
al. (1995)\nocite{qmlg95} and Johnston et al. (2008)\nocite{jkmg08},
are summarized in Table \ref{highB}. Apart from the high $B_s$, there
is nothing exceptional about RRAT J1819$-$1458 or any of the listed
pulsars in this context. Despite the large values of $B_s$, there are
pulsars with high, moderate or low polarization in this sample. This
classification is based on the peak degree of linear polarization in
at least one profile component, with the thresholds for high and
moderate polarization defined at $\approx$75\% and $\approx$40\%
respectively. As expected, the highly polarized sources are associated
with the highest values of $\dot E$.
\begin{table}
  \caption{\label{highB} The polarization (at 1.4~GHz) and $\dot E$
    of  RRAT J1819$-$1458 and 10 high
    $B_s$ pulsars from Gould \& Lyne (1998), D'Allessandro \& McCulloch
    (1997), Qiao et al. (1995) and Johnston et al. (2008).}  
\centering 
\small
\begin{minipage}{200mm}
\begin{tabular}{@{}lccl} 
  \hline 
  PSR  &  $B_s \times 10^{13}$~Gauss  &  $\dot E$ erg~s$^{-1}$ &   \% polarization  \\ 
  \hline 
  J1819$-$1458 & 4.96 & $2.88 \times 10^{32}$ & low \\
  B0154+61   &    2.13 &  $5.73 \times 10^{32}$ & low \\
  B1740$-$31    &   1.73 &  $3.39 \times 10^{32}$ & low \\
  B1737$-$30    &   1.70 &  $8.24 \times 10^{34}$ & high \\
  B1916+14   &    1.60 &  $5.09 \times 10^{33}$ & moderate \\
  B1849+00   &    1.47 &   $3.69 \times 10^{32}$ & low \\
  B2002+31   &    1.27 &  $3.13 \times 10^{32}$ & low \\
  B0525+21   &    1.24 &  $3.01 \times 10^{31}$ & moderate \\
  B1727$-$47    &   1.18 &  $1.13 \times 10^{34}$ & low \\
  B1610$-$50    &   1.08 &  $1.57 \times 10^{36}$ & high \\
  B1845$-$19    &   1.01 &  $1.15 \times 10^{31}$ & low \\
  \hline 
\end{tabular} 
\end{minipage}
\end{table}

The radio emitting magnetar XTE J1810$-$197 makes for an intriguing
counter-example to the high polarization - high $\dot E$ link. It is
100\% polarized, with an $\dot E$ value of only $1.8 \times
10^{33}$~erg s$^{-1}$(Camilo et al. 2006)\nocite{crh+06}, indicating a
dependence of the degree of polarization on some other parameter, to
which $\dot E$ may also be linked (e.g. the emission height,
Karastergiou \& Johnston 2007)\nocite{kj07}. This provides an
interesting context for a proposition regarding the nature of RRAT
J1819$-$1458, originally put forward by McLaughlin et
al. (2007)\nocite{mrg+07}. By scrutinizing the X-ray spectrum of the
source, the authors noticed the comparable temperature of the soft
X-ray spectrum to the quiescent state of the magnetar XTE
J1810$-$197. They therefore considered the possibility that RRAT
J1819$-$1458 represents a transitional class of object, between the
pulsar and magnetar populations. The polarization data presented here
does not strengthen or weaken this scenario, but points towards an
observational test. As XTE 1810$-$197 is gradually returning to
quiescence, it will be as interesting to look for orthogonal PA jumps
and a decrease in the linear polarization, as for intermittent
emission typical of RRATs (tentative evidence of the latter can be
found in observations in Hotan et al. 2007\nocite{hldd07}).

\section{Conclusions}
We have shown the first polarization measurements of the
RRAT~J1819$-$1458. The integrated linear polarization is moderate to
low, owing to the presence of competing orthogonal modes of
polarization, although highly polarized individual pulses are
observed. The polarization PA traces an S-shaped curve, interrupted by
2 orthogonal jumps, in agreement with the . This confirms both the dipolar structure of the
magnetic field at the region where the polarization is set, and the
three component pulse profile which leads to the observed bands in the
timing residuals. The degree of polarization is low, in accordance
with the measured $\dot E$ and despite the young age, strengthening
the observational link between $\dot E$ and the degree of
polarization.  Polarization observations at higher and lower
frequencies should lead to a more complete comparison with the
emission properties of PSR B0656+14. If the polarization of
RRAT~J1819$-$1458 behaves like normal pulsars, it should appear more
polarized at low frequencies, and probably completely unpolarized at
higher frequencies. Finally, it will be interesting to monitor the
polarization properties of the radio emitting magnetars to search for
orthogonal PA transitions as these sources return to their quiescent
state, thereby testing the possibility that RRAT J1819$-$1458
represents a transition between pulsars and magnetars.
\section*{Acknowledgments}
We thank John Sarkissian and John Reynolds, for assistance with the
observations and Joanna Rankin for a constructive referee's report. We
are also grateful to all the astronomers present at the telescope who
were kind enough to lend a hand in recording our data. The Australia
Telescope is funded by the Commonwealth of Australia for operation as
a National Facility managed by the CSIRO. AK is indebted to the
Leverhulme Trust for financial support. MAM is supported by a WV
EPSCOR Research Challenge Grant.

\bibliography{journals,modrefs,psrrefs,crossrefs,somerefs}
\bibliographystyle{mn2e}
\label{lastpage}

\end{document}